\documentclass[a4paper,11pt]{article}
\pdfoutput=1
\usepackage{jheppub}
\usepackage[utf8]{inputenc}
\usepackage{amsmath}
\usepackage{amssymb}
\title{\boldmath Entanglement Entropy for D3-, M2- and M5-brane backgrounds}

\author{Edward Quijada and Henrique Boschi-Filho}

\affiliation{Instituto de F\'\i sica, Universidade Federal de Rio de Janeiro,\\Caixa Postal 68528, RJ 21941-972, Brazil}

\emailAdd{edward@if.ufrj.br}
\emailAdd{boschi@if.ufrj.br}

\abstract{In the context of the gauge/gravity duality applications, we  study and compute the 
entanglement entropy of gauge theories  corresponding to  string/M theories on D3-, M2- and M5-brane  backgrounds. 
This is achieved using the Ryu-Takayanagi formula. We obtain the entanglement entropy for the general cases of the D3-, M2- and M5-brane 
backgrounds and also for the near horizon AdS limit, as well as the non-conformal flat space limit in each case. } 

\begin{document}
\maketitle
\flushbottom

\section{Introduction}

The thermal entropy of black holes obey an area-law established by Bekenstein \cite{Bekenstein:1973ur}
and Hawking  \cite{Hawking:1976de} and it was re-obtained using string theory in \cite{Strominger:1996sh}. 
The relation between entropy and temperature of black holes and string theory were further investigated in \cite{Gubser:1996de} where the entropy and
temperature of non-extremal black 3-branes were calculated. 

Soon after, Maldacena proposed the AdS/CFT correspondence \cite{maldacena1} conjecturing a duality between string/M-theory in anti-de Sitter space in $(d+2)$ 
dimensions times a compact space $\cal M$ with supercorformal field theories in $(d+1)$-dimensional flat spacetime. 
The black hole entropy in de Sitter space as well as the quantum entanglement were then analysed using the AdS/CFT correspondence in \cite{Hawking:2000da}.
In that work, it was found that the quantum entanglement entropy can also be viewed as the entropy of the thermal Rindler particles near the horizon, 
thereby avoiding reference to the unobservable region behind the horizon, which is the usual situation of the entanglement entropy. 

A major advance in the calculation of the entanglement entropy was achieved by Ryu-Takayanagi (RT) \cite{Ryu:2006bv}. As we know,
one can define the entanglement entropy $S_A$ in a gauge theory on $\mathbb{R}^{(1,d)}$ for a subsystem $A$ that has an arbitrary $(d-1)$-dimensional 
boundary $\partial A \subset \mathbb{R}^d$. In this set up, the proposal for the entanglement entropy is given by:
\begin{equation}\label{s0}
 S_A=\frac{\text{Area}(\gamma_A)}{4G_N^{(d+2)}},
\end{equation}
where
$\gamma_A$ denotes the $d$ dimensional static minimal surface in $AdS_{d+2}$ whose boundary coincides 
with the boundary of region $A$ ($\partial A=\partial\gamma_A$) and $G_N^{(d+2)}$ is the $(d+2)-$dimensional Newton's constant. 

Further developments on the entanglement entropy including the application to non-conformal cases were discussed in \cite{Ryu:2006ef}. 
Specifically, those non-conformal cases considered in \cite{Ryu:2006ef} were the ones corresponding to  dilatonic  D2- and NS5-branes as the gravity 
backgrounds. In our paper we work with non-dilatonic extremal brane solutions and  we will only obtain non-conformal cases when we go beyond the strong 
coupling limit. 

With the aim of  understanding the time-dependence of the entanglement entropy for generic quantum field theories,  a covariant holographic entanglement entropy 
proposal was presented in \cite{Hubeny:2007xt}. 
In ref. \cite{Faraggi:2007fu}  the holographic entanglement entropy and phase transitions at finite temperature were studied. 

Geometric entropy, which is related to entanglement entropy with a double Wick rotation, was introduced in  \cite{Fujita:2008zv} as a parameter of order in 
confinement/deconfinement transitions, as expected, this entropy becomes discontinuous at 
the Hagedorn temperature  both in free $\mathcal{N}=4$ super Yang-Mills, and its supergravity duals.

As opposed to the thermal entropy, the entanglement entropy is non-vanishing at zero temperature. Therefore we can employ it to probe the quantum properties of
the ground state for a given quantum system \cite{Nishioka:2009un}.

With the motivation of quantifying the degree of superhorizon correlations that are generated by the cosmological expansion, a computation of entanglement entropy for
quantum field theories in de Sitter space was presented 
in \cite{Maldacena:2012xp}.  
Other authors \cite{Bea:2015fja} have calculated  entanglement entropy for other CFTs that were  obtained as a result of  finding new compactification 
spaces in the dual string theory on $AdS_3\times\mathcal{M}$. This entanglement entropy were considered as a way of characterizing these new CFTs . 
A proposal on how to derive properties of the bulk geometry from the starting point of abstract quantum states in a Hilbert space using the entanglement entropy was 
presented  in  \cite{Cao:2016mst}. 
In refs. \cite{Mishra:2015cpa, Mishra:2016yor, Ghosh:2017ygi} 
the  holographic entanglement entropy has been calculated for
the boosted blackbrane  up to second order.

In our paper we  compute and analyze the entanglement entropy of large $N$  gauge theories holographically dual to string/M-theories on D3-, M2- and M5-brane supergravity backgrounds. 
First, we derive analytically this entropy for the limit geometries  $AdS_5\times S^5$, $AdS_4\times S^7$
and $AdS_7\times S^4$ of the brane cases, confirming the results obtained in \cite{Ryu:2006bv}, and also  for the  Minkowski-like limit geometries of  these same brane background spaces. The case of the D3-brane was discussed recently in connection with the open-closed string duality \cite{Niarchos:2017cdz}. 

As it is know, the curvature of $AdS$ spaces are constant. On the contrary, the curvature of the general non-dilatonic extremal solutions of supergravity, i.e  
D3-, M2- and M5-brane background spaces,   are not constant.  So  it would be interesting to analyze how possible geometric transitions (i.e. at zero 
temperature) could occur when one 
goes from one geometric regime to the other which are both contained in these background spaces. For such aim, in this work we also do a numerical study comparing the entanglement entropy
in some brane-background spaces with their asymptotic regimes.

This paper is organized as follows, in sec. 2 we study the entanglement entropy in the D3-brane background. We do the same in sec. 3 and sec. 4,  for M2- and M5-brane 
backgrounds, respectively. Finally, in sec. 5, we present some conclusions on this paper and some ideas for possible future works.


\section{Entanglement entropy for the  D3-brane background}\label{s1}
\begin{figure}[h]
 \centering
 \includegraphics[width=6cm,height=5cm]{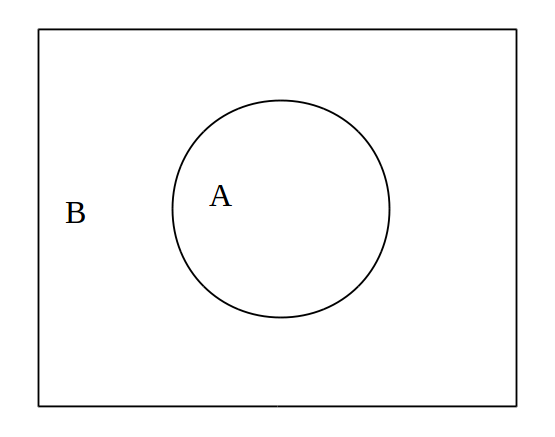}
 \caption{ Subsystems  $A$ and $B$ in the space of a field theory.}
 \label{subsistem}
\end{figure}

Consider a quantum field theory or a many-body system defined on $(d+1)$-dimensional region $A \cup B$ ($A\cap B=\emptyset$), see fig. \ref{subsistem}, where $A$ and $B$ are $d$-dimensional space-like manifolds.
Then, the entanglement entropy  $S_A $ is defined as the Von Neumann entropy $S_A$ of the reduced density matrix when we take the partial trace with respect to
the degrees of freedom inside $B$. 
The entanglement entropy $S_A$ measures how the subsystems $A$ and $B$ are correlated with each other. In other words,  this is the entropy for an observer in $A$
who is not accessible to $B$. 

From the point of view of holography, this entanglement entropy can be seen as a geometric object in the bulk. This object is  the minimal area surface
in the bulk that has the same boundary of that of the subsystem $A$. 

As a first example we do the computation of the entanglement entropy  ($S_A$) corresponding to  the space-time generated by a large 
$N$ number of coincident D3-branes. The invariant measure for this geometry is: \cite{Horowitz:1991cd,Gubser:1998bc}
\begin{equation}\label{0}
 ds^2=dx^{\mu}dx^{\nu}g_{\mu\nu} =\left(1+\frac{R^4}{r^4}\right)^{-1/2}(-dt^2 + dx_3^2) +\left(1+\frac{R^4}{r^4}\right)^{1/2}(dr^2 + r^2 d\Omega^2_5),
\end{equation}
where $x^{\mu}=(t,x^a,r,\theta_b )$, with $a=1,2,3$ and $b=1,2...5$, are the space-time coordinates, $g_{\mu\nu}$ 
is the space-time metric,  $R =(4\pi g N l_s^4)^{1/4}$, $dx_3^2$ is the euclidean line element in $\mathbb{R}^3$ and $d\Omega^2_5$ is the line element
on the 5-sphere which only depends on $\theta_b$ coordinates.


\subsection{Rectangular strip}
Throughout  this paper  we  choose  an infinite rectangular strip as the subsystem $A$. This rectangular strip will be defined for each background geometry in the next  sections. 
The style of the computations performed in this paper were inspired in those that can be found in  \cite{Fischler:2013gsa}.

For a 3-dimensional ($d=3$) subsystem $A\in \mathbb{R}^3$ we choose the following rectangular strip region:
\begin{equation}\label{regionx}
 X\equiv x^1\in\left[-\frac{l}{2},\frac{l}{2}\right]\,\,\,,\,\,\, x^2,x^3\in\left[-\frac{L}{2},\frac{L}{2}\right].
\end{equation}

The corresponding extremal surface is translationally invariant along 
$x^2$, $x^3$ and the
profile in the bulk is given by $x^1\equiv X(r)$. The  area of this  surface is  given by
\begin{equation}
 \mathcal{A}=\int d^3\sigma\sqrt{det(G_{\alpha\beta})},
\end{equation}
where $G_{\alpha\beta}\equiv \partial_{\alpha}x^{\mu}\partial_{\beta}x^\nu g_{\mu\nu}$ is the induced metric on the 
surface $\gamma_A$ in the bulk. We took the parametrization of
$\gamma_A$ as : $\sigma_1=r$, $\sigma_2=x^2 $, $\sigma_3=x^3$. As a result we have:
\begin{equation}\label{7.1}
 \mathcal{A}=L^2\int \frac{dr}{f(r)}\sqrt{\frac{X'^2}{f(r)}+f(r)},
\end{equation}
where $f(r)=\sqrt{1+R^4/r^4}$, and $X'\equiv dX/dr$.

\begin{figure}
 \centering
 \includegraphics[width=8cm, height=6cm]{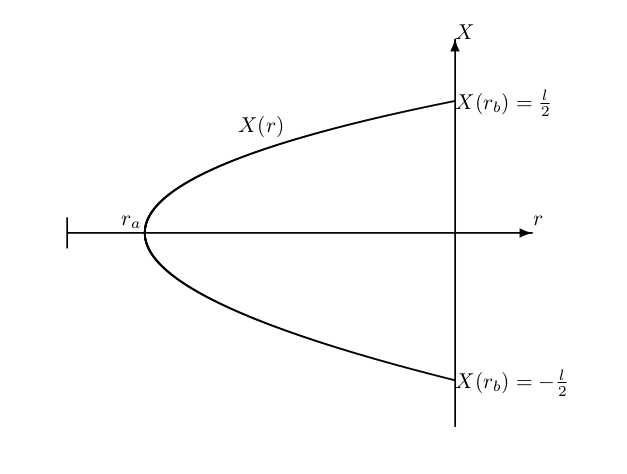}
\caption{Profile $X(r)$ of the $\gamma_A $ surface with boundary conditions.}
\label{figs}
\end{figure}

From the functional area (\ref{7.1}) we obtain the equation for the profile $X(r)$ that makes the area $A$ extremal: 
\begin{equation}\label{7.22}
 \frac{dX}{dr}=\pm\frac{f^{5/2}(r)}{\sqrt{f^3(r_a)-f^3(r)}},
\end{equation}
where that $r_a$ is the closest approach of the extremal surface (see fig.\ref{figs}) to the origin of $r$ coordinate.
Such  surface has two branches, joined smoothly at $r=r_a$, where $X=0$ and $X'\rightarrow\infty$. Additionally we choose an UV cut-off 
value $r_b$ that defines the next boundary condition for the profile $X(r)$:
\begin{equation}\label{UV}
 X(r_b)=\pm\frac{l}{2},
\end{equation}
With this condition  we integrate (\ref{7.22})  to obtain the width of the strip:
\begin{equation}\label{ld3}
 \frac{l}{2}=\int_{r_a}^{r_b}dr\frac{f^{5/2}(r)}{\sqrt{f^3(r_a)-f^3(r)}}.
\end{equation}
Note  that $r_a$ can be determined in terms of $l$ and $r_b$ from this last equation.
After substitution of eq.(\ref{7.22}) into eq.(\ref{7.1}) and considering the limits of integration defined in eq.(\ref{ld3})  we  finally get the extremal area: 
\begin{equation}\label{areafun}
 \mathcal{A}=L^2\int_{r_a}^{r_b}dr\frac{f^{3/2}(r_a)}{\sqrt{f(r)[f^3(r_a)-f^3(r)]}}.
\end{equation}
Since $f(r)$ is a monotonic decreasing function of $r$ with maximum $f(r_a)$ and minimum $f(\infty)\rightarrow 1$,
the expression for the area $\mathcal{A}$  diverges as $r_b\rightarrow\infty$ because the integrand of $\mathcal{A}$ goes to a constant factor
as  $r\rightarrow\infty$.
 So in order to obtain a finite result we need 
to separate the divergent part from the extremal area $\mathcal{A}$ as follows:
 \begin{equation}
  \mathcal{A}=\mathcal{A}_{\text{div}}+\mathcal{A}_{\text{finite}}.
 \end{equation}
The divergent part can be defined from (\ref{areafun}) as:
 \begin{equation}\label{Area-div}
\mathcal{A}_{\text{div}}=L^2\int_{0}^{r_b}dr\frac{f^{3/2}(r_a)}{\sqrt{f(r)[f^3(r_a)-1]}}.
 \end{equation}
Then, using the RT-formula (\ref{s0}) we may schematically  write the entanglement entropy as:
\begin{equation}\label{entropy}
 S_A=S_{\text{div}}+\frac{L^2R}{4G^5_N}\,\,s\left(\frac{l}{R}\right),
\end{equation}
where $s(l/R)$ is the finite dimensionless entanglement entropy. This function is an implicit function, this means that it is not always possible to 
find it explicitly as an algebraic formula in terms of $l$. Instead in general we
 only are able to write the expression for $s$ in terms of $r_a$ and $r_b$. 
\begin{equation}
 s(r_a)=\int_{r_a}^{r_b}\frac{dr}{R}\frac{f^{3/2}(r_a)}{\sqrt{f(r)[f^3(r_a)-f^3(r)]}}-\int_{0}^{r_b}\frac{dr}{R}\frac{f^{3/2}(r_a)}{\sqrt{f(r)[f^3(r_a)-1]}}.
 \end{equation}


\subsection{Near horizon limit: $AdS_5\times S^5$}

This limit geometry of the N D3-branes is obtained when $R>>r$, so that 
in this regime $f(r)\approx \frac{R^2}{r^2}$.
This asymptotic limit corresponds to the $\mathcal{N}=4$ superconformal field theory in $D=4$ and gauge group $SU(N)$.
Throughout all this paper  we are considering the limit  $r_a<<r_b$ so we have that the width of the strip (\ref{ld3}) in this regime is:
\begin{equation}\label{123}
\frac{l}{R}=\frac{2R}{r_a}\int_{1}^{\frac{r_b}{r_a}}\frac{dx}{x^2\sqrt{x^6-1}}= \frac{R}{r_a}c\left[1+\mathcal{O}\left(\frac{r_a^4}{r_b^4}\right)\right],
\end{equation}
where $c=\frac{2\sqrt{\pi}\Gamma(2/3)}{\Gamma(1/6)}$. From now on we are going to disregard the superior order $\mathcal{O}\left(\frac{r_a^4}{r_b^4}\right)$ in our
computations.

Furthermore, from eq.(\ref{areafun}) and (\ref{s0}) we have that the entanglement entropy in this regime is:
\begin{equation}\label{sadsra}
 S_A=\frac{L^2r_a^2}{4G_N^5R}\int_1^{\frac{r_b}{r_a}}dx\frac{x^4}{\sqrt{x^6-1}}=\frac{L^2r_b^2}{24G_N^5R}-\frac{L^2cr_a^2}{16G_N^5R}.
\end{equation}

From this last equation and eq.(\ref{123})   we can  eliminate the parameter $r_a$ in order to  have  the  entanglement entropy in terms  of $l$ and $r_b$: 
\begin{equation}\label{near-en}
S_A=\frac{r_b^2L^2}{8G_N^5R}-\frac{L^2c^3R^3}{16G_N^5l^2}.
\end{equation}
The first term of this equation is divergent as $r_b\rightarrow\infty$ in accordance with eq.(\ref{Area-div}). This result  
agrees with the one obtained in 
\cite{Ryu:2006bv}
up to an arbitrary constant factor in the divergent part of the entanglement entropy. Then the  dimensionless finite entanglement entropy is in this case:
\begin{equation}
 s(l/R)=-\frac{c^3R^2}{108 l^2}.
\end{equation}


\subsection{Flat-space limit}

This case corresponds to an almost flat-space geometry which is achieved when we take the limit $R<<r$ which implies that $f(r)\approx1+\frac{R^4}{2r^4}$.
The corresponding dual field theory in this case (when it exist) would be 
 $SU(N)$ gauge theory in $D=4$, non-conformal nor supersymmetric any more.  
In this limit  the width of the strip (\ref{ld3}) results:
\begin{equation}\label{flatl1}
 \frac{l}{R}\approx\frac{2\sqrt{2/3}}{R^2}r_a^3\int_1^{\frac{r_b}{r_a}}dx\frac{x^2}{\sqrt{x^4-1}}\approx2\sqrt{2/3}\left(\frac{r_a}{R}\right)^2\frac{r_b}{R}.
\end{equation}
As we can see this expressions  for $l/R$  diverges as $r_b\rightarrow\infty$.

Furthermore, from eqs. (\ref{s0}) and (\ref{areafun}) we have that the entanglement entropy  in this regime  is:
\begin{equation}\label{eeflat}
S_A \approx\frac{L^2}{4G_N^5}\frac{\sqrt{2/3}}{R^2}r_a^3\int_1^{\frac{r_b}{r_a}}dx\frac{x^2}{\sqrt{x^4-1}}
=\frac{\sqrt{2/3}}{16G_N^5}\frac{L^2r_a^2r_b}{R^2}-\frac{c^*\sqrt{2/3}}{4G_N^5}\frac{L^2r_a^3}{R^2},
\end{equation}
where $c^{*}=\frac{\sqrt{\pi}\Gamma(3/4)}{\Gamma(1/4)}$. This expression for the entanglement entropy naturally contains a term that is divergent as 
$r_b\rightarrow\infty$, this term is in accordance with eq.(\ref{Area-div}).
Then from eqs. (\ref{flatl1}) and (\ref{eeflat}) we can eliminate the parameter $r_a$ in order to
have the entanglement entropy in terms of  $l$ and $r_b$:
\begin{equation}\label{sad3}
S_A=\frac{L^2l}{32G_N^5R}-\frac{c^*(3/8)^{\frac{1}{4}}L^2R}{8G_N^5}\left(\frac{l}{r_b}\right)^{3/2}.
\end{equation}
The first term of this expression correspond to the divergent part of the entropy, since as we can see from eq.(\ref{flatl1}) $l\sim r_b$.
Then the dimensionless finite entanglement entropy is:
\begin{equation}
 s(l/R)=-\frac{c^*(3/8)^{\frac{1}{4}}}{2}\left(\frac{l}{r_b}\right)^{3/2}.
\end{equation}
Note that this novel result still contains a dependence on $r_b$. This is not a problem because this expression remains finite as $r_b\rightarrow\infty$.
This follows  from eq.(\ref{flatl1}), where as one can see the coefficient $l/r_b$ remains finite 
as $r_b\rightarrow\infty$.


\subsection{D3-brane and  asymptotic limits}

\begin{figure}[h]
 \centering
 \includegraphics[width=10cm,height=6cm]{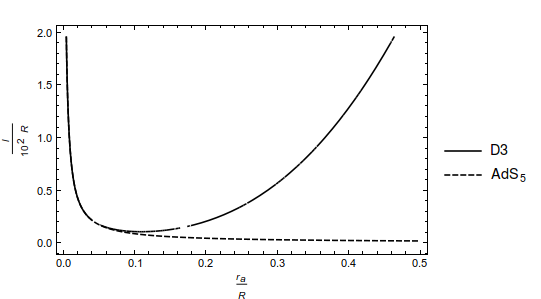}
 \caption{Width of the strip $l/R$ against $r_a/R$ for $r_b/R=10^3$ and $r_a/R<0.4$.} 
 \label{fig9}
\end{figure}

In the case of the  D3-brane background space we make a numerical analysis in order to know the behaviour of the entanglement entropy. 
Then in fig.\ref{fig9}  we present a numerical plot of  $l/R$ in eq.(\ref{ld3}) against $r_a/R$.
Actually this plots shows the comparison between the AdS limit and the D3-brane space, the agreement is conspicuous  in the region where $r_a/R<0.1$, which
is the  minimum of the D3-brane plot. At this minimum the width of the strip  is $l/R\approx 10.4$ for the D3-brane case. In this example we have chosen $r_b/R=10^3$ 
as the cut-off. See also \cite{Niarchos:2017cdz} for a recent related discussion. 

\begin{figure}[h]
 \centering
 \includegraphics[width=9cm,height=6cm]{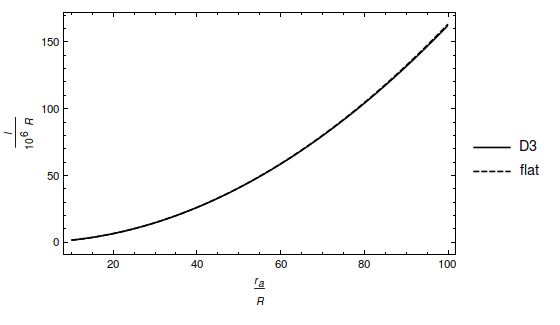}
 \caption{ Width of the strip $l/R$ against $r_a/R$  for $r_b/R=10^4$ and $r_a/R>10$.} 
 \label{lfd3}
\end{figure}

The next plot we present in fig.\ref{lfd3} shows the comparison between the D3-brane case and the flat-space approximation for the
width of the strip $l/R$ against $r_a/R$. In this case a cut-off of $r_b/R=10^4$ was used and we considered a region $r_a/R>10$. As we can see from this 
plot, the agreement between both backgrounds are satisfactory in that region.

\begin{figure}[h]
 \centering
 \includegraphics[width=9cm,height=7cm]{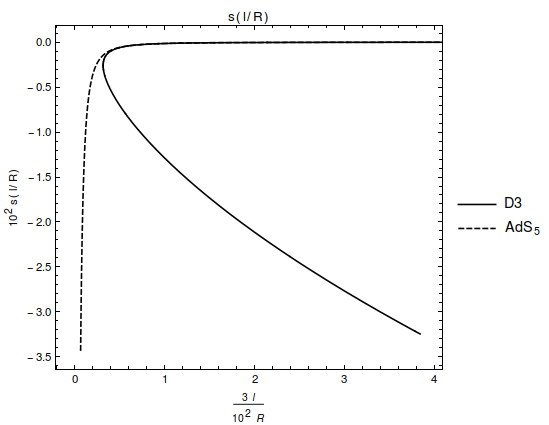}
 \caption{Entanglement entropy $s$  against $l/R$ for $r_b/R=10^3$ and $r_a/R<0.4$.} 
 \label{fig10}
\end{figure}

In the next plot of fig.\ref{fig10} this time we present the  dimensionless entanglement entropy $s$ against $l/R$. Actually a comparison between the 
plots  for the D3-brane and AdS backgrounds are shown. This  was made for the  cut-off  of $r_b/R=10^3$ in the region $r_a/R<0.4$.
From this plot we can see that for the D3-brane background the entanglement entropy $s$  has two 
branches. 
The superior branch corresponds to the region $r_a/R<0.1$
and the inferior one to region $r_a/R>0.1$. Note that the superior branch coincides with  the AdS case.

\begin{figure}[h]
 \centering
 \includegraphics[width=10cm,height=7cm]{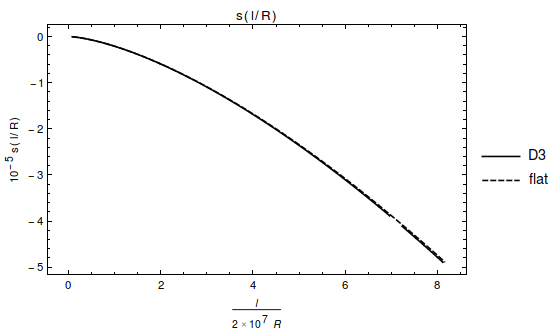}
 \caption{Entanglement entropy $s$ against for $r_b/R=10^4$ and $r_a/R>10$.} 
 \label{fig13}
 \end{figure}

In the next plot of fig.\ref{fig13} we present a comparison between the entanglement entropy of the D3-brane and flat-space backgrounds. In this plot 
we consider the cut-off of $r_b/R=10^4$ and the region $r_a/R>10$. We see from this figure that the plot of both backgrounds have a satisfactory agreement in that region.


\section{Entanglement entropy for the M2-brane background}

In this section we do the computation of entanglement entropy ($S_A$) corresponding to 
the 11-dimensional supergravity M2-brane background. The metric solution generated by N coincident M2-branes is \cite{Review2}: 
\begin{equation}
ds^2_{\text{M2}}=dx^{\mu}dx^{\nu}g_{\mu\nu}=\left(1+\frac{R_2^6}{r^6}\right)^{-2/3}(-dt^2+ dx_2^2)+\left(1+\frac{R_2^6}{r^6}\right)^{1/3}(dr^2+r^2d\Omega^2_7), 
\end{equation}
where $x^{\mu}=(t,x^a,r,\theta_b )$, with $a=1,2$ and $b=1,2...7$, are the space-time coordinates, $g_{\mu\nu}$ 
is the space-time metric,  $R_2=(32\pi Nl^6_{11})^{1/6}$, $l_{11}$ is the Plank's length in eleven dimensions, $dx_2^2$ is the euclidean line element in $\mathbb{R}^2$ and $d\Omega^2_7$ is the line element
on the 7-sphere which only depends on $\theta_b$ coordinates.


\subsection{Rectangular strip}

As we worked in the last section, we take  a 2-dimensional ($d=2$) subsystem  $A\in \mathbb{R}^2$ as  the following rectangular strip:
\begin{equation}
X=x^1\in\left[-\frac{l}{2},\frac{l}{2}\right]\,\,\,\,,\,\,\,\,\,x^2\in\left[-\frac{L}{2},\frac{L}{2}\right]
\end{equation}

Now we parametrize the $\gamma$-surface as follows: $x^1=X(r)$, $\sigma_1=r$ and $\sigma_2=x^2$. The rest of the coordinates are independent
of this parametrization.
Then the area of this surface
is given by:
\begin{equation}
\mathcal{A}=\int d^2\sigma\sqrt{det(G_{\alpha\beta})},
 \end{equation}
where $G_{\alpha\beta}=\partial_{\alpha}x^{\mu}\partial_{\beta}x^{\nu}g_{\mu\nu}$ is the induced metric on the $\gamma$-surface.

After the computation of   this area $\mathcal{A}$ in the M2-brane background we obtain:
\begin{equation}\label{A1}
 \mathcal{A}=L\int dr\sqrt{\frac{X'^2}{f^4(r)}+\frac{1}{f(r)}}
\end{equation}
where $f(r)=\left(1+\frac{R^6_2}{r^6}\right)^{1/3}$. 

From this area functional  we can derive  the equation for the profile $X(r)$ of the surface that extremize this area, this equation is
given by: 
\begin{equation}\label{eq1}
 \frac{dX}{dr}=\pm\sqrt{\frac{f^7(r)}{f^4(r_a)-f^4(r)}}.
\end{equation}
As in the previous section  $r_a$ represents
the closest approach of the profile $X(r)$ to the origin of coordinates. This profile has  two branches that join 
smoothly at $r_a $, where $X=0$ and $X'\rightarrow\infty$. We also take $r_b$ as the  UV cut-off in   such a way  that the width of the strip $l$ 
is given by:
\begin{equation}\label{x2}
X(r_b)=\pm\frac{l}{2}.
 \end{equation}
 If we solve the eq.(\ref{eq1}) with the boundary condition (\ref{x2}) we obtain the width of the strip as an integral expression that depends on
 $r_a$ and $r_b$.
\begin{equation}\label{l2}
 \frac{l}{2}=\int_{r_a}^{r_b}dr \sqrt{\frac{f^7(r)}{f^4(r_a)-f^4(r)}}.
\end{equation}
As was pointed out in the previous section  $r_a$ could be  determined by  inverting eq.(\ref{l2}). The result would be in terms of $l$ and $r_b$. 
After substitution  of the equation for the extremal surface (\ref{eq1}) into  the functional area (\ref{A1}) and 
taking the limits of integration used in eq.(\ref{l2}) we finally obtain the extremal area:
\begin{equation}\label{A2ex}
 \mathcal{A}=L\int_{r_a}^{r_b}dr\sqrt{\frac{f^4(r_a)}{f(r)[f^4(r_a)-f^4(r)]}}.
\end{equation}
As we can observe $f(r)$ is a monotonic decreasing function with maximum $f(r_a)$ and minimum  $f(\infty)\rightarrow 1$. Then the integrand of eq.(\ref{A2ex}) goes 
to a constant as $r\rightarrow\infty$. Hence we have that the area $\mathcal{A}$ is divergent as $r_b\rightarrow\infty$. The divergent part of the
area $\mathcal{A}$ can be defined in terms of UV cut-off $r_b$ as:
\begin{equation}\label{AdivM2}
 \mathcal{A}_{\text{div}}=L\int_{0}^{r_b}dr\sqrt{\frac{f^4(r_a)}{f(r)[f^4(r_a)-1]}}.
\end{equation}
The entanglement entropy can be computed from area $\mathcal{A}$ and  RT-formula (\ref{s0}). We may write schematically this entropy as 
follows: 
\begin{equation}\label{ee2}
 S_A=S_{\text{div}}+\frac{LR_2}{4G_N^4}\,\,s\left(\frac{l}{R_2}\right),
\end{equation}
where $S_{\text{div}}$ is the divergent part of the entropy and $s(l/R_2)$ is a dimensionless finite part of the entanglement entropy as
a function of $l/R_2$. Not always it is possible to put this dimensionless entropy as a explicitly function of $l/R_2$. In general 
what we can only do is to put this dimensionless entropy in terms of $r_a$ and $r_b$: 
\begin{equation}\label{eem2}
 s(r_a)=\int_{r_a}^{r_b}\frac{dr}{R_2}\sqrt{\frac{f^4(r_a)}{f(r)[f^4(r_a)-f^4(r)]}}-\int_{0}^{r_b}\frac{dr}{R_2}\sqrt{\frac{f^4(r_a)}{f(r)[f^4(r_a)-1]}}.
\end{equation}


\subsection{Near horizon limit: $AdS_4\times S^7$}

This limit geometry of the N M2-branes are obtained when we take  $R_2>>r$ so  that $f\approx\frac{R^2_2}{r^2}$.
In this case the dual theory corresponds to the ABMJ theory in $D=3$ 
\cite{Aharony:2008ug}. 
Then if we compute  the width of the strip (\ref{l2}) in this background  limit we obtain: 
\begin{equation}\label{1234}
\frac{l}{R_2}= \frac{2R_2^2}{r_a^2}\int_1^{\frac{r_b}{r_a}}\frac{dx}{x^3\sqrt{x^8-1}}=\frac{R_2^2}{r_a^2}c_2\left[1+\mathcal{O}\left(\frac{r_a^6}{r_b^6}
\right)\right],
\end{equation}
where $c_2=\frac{\sqrt{\pi}\Gamma(3/4)}{\Gamma(1/4)}$. In what follows we are going to disregard the superior order 
$\mathcal{O}\left(\frac{r_a^6}{r_b^6}\right)$ in this result. 

The entanglement entropy in this limit background can be calculated from the extremal area $\mathcal{A}$ (eq.\ref{A2ex}) and 
the RT-formula (eq.\ref{s0}), as a result we obtain:
\begin{equation}
 S_A=\frac{Lr_a^2}{4RG_N^4}\int_1^{r_b/r_a}dx\frac{x^5}{\sqrt{x^8-1}}=\frac{Lr_b^2}{4G_N^4r_a}-\frac{Lc_2r_a^2}{8G_N^4R_2}.
\end{equation}
Notice that the  first term of this expression is divergent as $r_b\rightarrow\infty$. This term can be also be obtained directly from
eq.(\ref{AdivM2}).

Now we eliminate the parameter $r_a$ from these last two equations in order to put the entanglement entropy $S_A$ as a function of the 
width of strip $l$:
\begin{equation}\label{ee2ads}
S_A=\frac{Lr_b^2}{4G_N^4R_2}-\frac{Lc_2^2R_2^2}{8G_N^4l}.
\end{equation}
Up to an arbitrary constant factor in the divergent part of this equation, this result agrees with the one obtained in \cite{Ryu:2006bv}.
Then in this regime the dimensionless finite entanglement entropy results:
\begin{equation}\label{eem2ads}
 s(l/R_2)=-\frac{c_2^2R_2}{2 l}.
\end{equation}


\subsection{Flat-space limit}

This background of an almost flat space is obtained as a result of assuming that  $R_2<<r$. 
This implies  that $f\approx 1+\frac{R_2^6}{3r^6}$.
This case corresponds, in the field theory side, to a non-conformal 
nor supersymmetric  gauge theory 
$SU(N)$ in $D=3$.
Then we can calculate  the width of the strip (eq.\ref{l2}):
\begin{equation}\label{lm2flat}
 \frac{l}{R_2}\approx\frac{\sqrt{3}r_a^4}{R_2^4}\int_1^{\frac{r_b}{r_a}}dx\frac{x^3}{\sqrt{x^6-1}} \approx\sqrt{3}\left(\frac{r_a}{R_2}\right)^3\frac{r_b}{R_2}.
\end{equation}
Also in this background limit we can calculate the entanglement entropy starting from eq.(\ref{A2ex}) and applying the RT-formula (eq.\ref{s0}):
\begin{equation}
 S_A\approx\frac{\sqrt{3}L}{8G_N^4R_2^3}r_a^4\int_1^{\frac{r_b}{r_a}}dx\frac{x^3}{\sqrt{x^6-1}}=
 \frac{\sqrt{3}Lr_a^3r_b}{48G_N^4R_2^3}-\frac{\sqrt{3}Lr_a^4c'}{8G_N^4R_2^3},
\end{equation}
where $ c'=\frac{\sqrt{\pi}\Gamma(5/6)}{\Gamma(1/3)}$. The first term of this expression  diverges as $r_b\rightarrow\infty$. This divergent term
can also be obtained directly from eq.(\ref{AdivM2}).

Now we  eliminate the parameter $r_a$ from these last two expressions in order to have  the entanglement entropy $S_A $ as a 
function of the width of the strip $l$:
\begin{equation}
 S_A=\frac{Ll}{48G_N^4}-\frac{Lc'R_2}{8(3)^{1/6}G_N^4}\left(\frac{l}{r_b}\right)^{4/3},
\end{equation}
As we can see the first term of this equation is proportional to $l$. This 
term corresponds to the divergent part of $S_A$. This is because from eq.(\ref{lm2flat}) the width of the strip $l\sim r_b$. 
Then the dimensionless entanglement entropy is:
\begin{equation}\label{ee2flat}
 s(l/R_2)=-\frac{c'}{2(3)^{1/6}}\left(\frac{l}{r_b}\right)^{4/3},
\end{equation}
This novel result  still contains the UV cut-off $r_b$, however, it continues to be a finite function. This is so because according to eq.(\ref{lm2flat}) 
$l/R_2$ is  finite and can be expressed in terms of $r_a$.  


\subsection{M2-brane and asymptotic limits}

\begin{figure}[h]
 \centering                 
 \includegraphics[width=10cm,height=6cm]{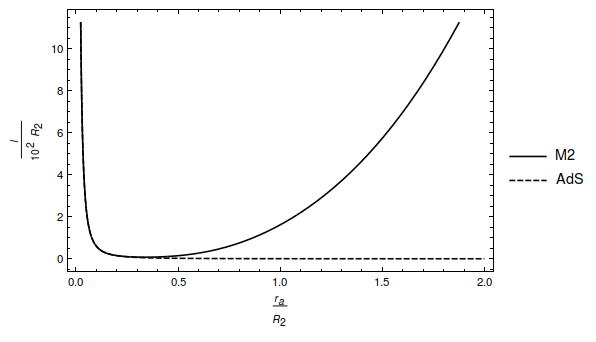}
 \caption{$l/R_2$ against $r_a/R_2$ for $r_b/R_2=10^2$.} 
 \label{fig21}
\end{figure}

In the case of the M2-brane background we can do a numerically study of the behaviour of the entanglement entropy. Let's start by 
 plotting (see fig.\ref{fig21}) a comparison between the width of the strip $l/R_2$ for the general M2-brane space (eq.\ref{l2}) and for the  AdS$_4$ 
 limit (eq.\ref{1234}). This  was done with  the cut-off  of $r_b/R_2=10^2$ and for the region $r_a/R_2<2$. As we can see from this figure both plots coincide
 for the region $r_a/R<0.32$, which is where the plot for the M2-brane has its minimum, which is  $l/R_2=8.32$ . 

\begin{figure}[h]
 \centering
 \includegraphics[width=10cm,height=6cm]{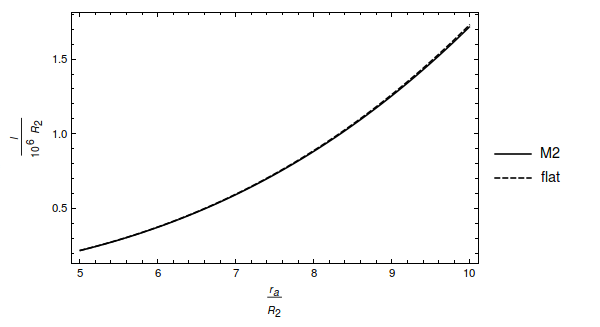}
 \caption{$l/R_2$ against $r_a/R_2$ for $r_b/R_2=10^3$.} 
 \label{fig22}
\end{figure}

Next we plot (see fig.\ref{fig22}) a comparison between the width of the strip for the M2-brane (eq.\ref{l2}) and  the almost-flat space  
limit (eq.\ref{lm2flat}).
This plot was done considering $r_b/R_2=10^3$ and  the region $r_a/R_2>5$. As we can see from this picture the coincidence is 
very good in such region.

\begin{figure}[h]
 \centering
\includegraphics[width=10cm,height=7cm]{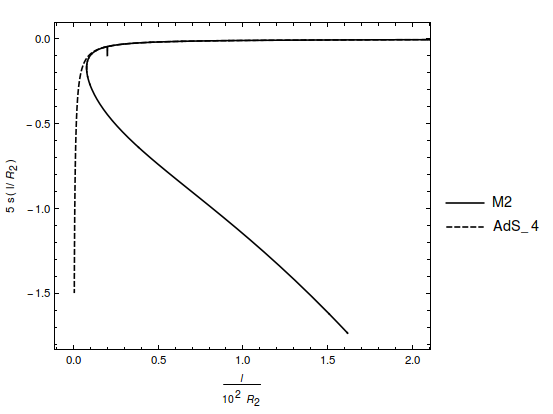}
 \caption{$s(l/R_2)$ against $l/R_2$ for $r_b/R_2=10^2$.} 
 \label{fig23}
\end{figure}

Now is time to analyse the dimensionless  entanglement entropy $s(l/R_2)$. So first we plot (see fig.\ref{fig23}) a comparison between the 
entanglement entropy for the general M2-brane space (eq.\ref{ee2}) and for the AdS$_4$ space limit (eq.\ref{eem2ads}). This plot was done considering 
the UV cut-off $r_b/R_2=10^2$ and the region $r_a/R_2<1$. As we can see from this figure there 
are two branches for the M2-brane background, the superior one corresponds to $r_a/R_2<0.32$ and the inferior one $r_a/R_2>0.32$.
The coincidence of the plots occurs in the superior branch of the M2-brane entropy plot.

\begin{figure}[h]
\centering
 \includegraphics[width=10cm,height=6cm]{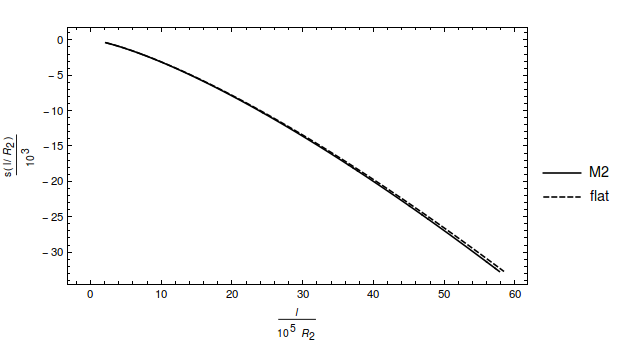}
 \caption{Entanglement entropy $s(l/R_2)$  for $r_b/R_2=10^3$.} 
 \label{fig24}
\end{figure}

Next we plot (see fig.\ref{fig24}) a comparison between the dimensionless entanglement entropy for the general M2-brane background (eq.\ref{ee2}) and 
for the almost-flat space limit ( eq.\ref{ee2flat}). In this case we used a UV cut-off $r_b/R_2=10^3$ and consider the region $r_a/R_2>5$. As we can see
from this figure, both plots coincide very well in such region. 


\section{Entanglement entropy for the M5-brane background}

In this section we  compute the entanglement entropy ($S_{\mathcal{A}}$) corresponding to the background space-time generated by
N coincident M5-branes. The 11-dimensional supergravity solution of M5-branes is given by the metric \cite{Review2,Review1}:
\begin{equation}\label{m5}
 ds^2_{\text{M5}}=g_{\mu\nu}dx^{\mu}dx^{\nu}=\left(1+\frac{R_5^3}{r^3}\right)^{-1/3}dx_6^2+\left(1+\frac{R_5^3}{r^3}\right)^{2/3}(dr^2+r^2d\Omega^2_4),
\end{equation}
where $x^{\mu}=(t,x^a,r,\theta_b )$, with $a=1,2,...5$ and $b=1,2...4$, are the space-time coordinates, $g_{\mu\nu}$ 
is the space-time metric,  $R_5=(\pi Nl_{11}^3)^{1/3}$ , $l_{11}$ is the Plank's length in eleven dimensions, $dx_5^2$ is the euclidean line element
in $\mathbb{R}^5$ and $d\Omega^2_4$ is the line element on the 4-sphere which only depends on $\theta_b$ coordinates.


\subsection{Rectangular strip}

In order to compute the entanglement entropy  we consider  a subsystem $A\in \mathbb{R}^5$ ($d=5$) as the following strip:
\begin{equation}\label{r1}
X\equiv x^1=\left[-\frac{l}{2},\frac{l}{2}\right]\,\,\,\,\,,\,\,\,\,\,\,x^{2,3,4,5}=\left[-\frac{L}{2},\frac{L}{2}\right]
\end{equation}

Now we parametrize the $\gamma$-surface as:  $x^1=X(r)$, $r=\sigma^1 $, $x^i=\sigma^i$ with $i=2,..5$. The rest of the coordinates are independent of
this prametrization.
 As we did in the previous sections we can compute  the area of the $\gamma$-surface in this parametrization following the next formula:
 \begin{equation}
  \mathcal{A}=\int d^5\sigma\sqrt{det(G_{\alpha\beta})},
 \end{equation}
where $ G_{\alpha\beta}=\partial x^{\mu}\partial x^{\nu}g_{\mu\nu}$ is the induced metric on the $\gamma$-surface.
Then the   area  $\mathcal{A}$ turns out to be  a functional of the surface profile $X(r)$ and can be written down as follows:
\begin{equation}\label{A3}
 \mathcal{A}=L^4\int dr\sqrt{\frac{X'^2}{f^5}+\frac{1}{f^2}},
\end{equation}
where $f=\left(1+\frac{R^3_5}{r^3}\right)^{1/3}$ and $X'\equiv dX/dr$.
From this functional  $\mathcal{A}$  we  can derive the equation  for the profile $X(r)$ that extremize the area of the surface. Such equation 
can be written down as  follows:
\begin{equation}\label{q2}
 \frac{dX}{dr}=\pm\frac{f^4(r)}{\sqrt{f^5(r_a)-f^5(r)}},
\end{equation}
where as in the previous sections $r_a$ is the closest point to the origin of coordinates, besides  $X(r_a)=0$ and
 $X'(r_a)\rightarrow\infty$. We also need to  define a  UV cut-off which we call $r_b$ and that satisfy the following boundary condition:
\begin{equation}\label{cut-off}
 X(r_b)=\pm\frac{l}{2}.
\end{equation}

Now we can compute the width of the strip $l$ in terms of the parameter $r_a$ and $r_b$. To do so we integrate eq.(\ref{q2}) subject to the
above boundary condition. Then the result is as follows:
\begin{equation}\label{l5}
 \frac{l}{2}=\int_{r_a}^{r_b}dr\frac{f^4(r)}{\sqrt{f^5(r_a)-f^5(r)}}.
\end{equation}
Note that we could obtain the parameter $r_a$ after inverting this equation. The result would be in terms of $l$ and $r_b$.

Now we will find the entanglement entropy $S_A$. In order to do so first we are going to find the extreme area $\mathcal{A}$ which follows 
from substituting the eq.(\ref{q2}) into the functional area (eq.\ref{A3}). The result is the next extremal area: 
\begin{equation}\label{m5-a}
\mathcal{A}=L^4\int_{r_a}^{r_b}dr\frac{1}{f(r)}\sqrt{\frac{f^5(r_a)}{f^5(r_a)-f^5(r)}}. 
\end{equation}
Since $f(r)$ is a monotonic decreasing function that has  maximum $f(r_a)$ and  minimum $f(\infty)=1$, we have that the integrand of this expression 
goes to a constant factor as $r\rightarrow\infty$. As a result we can conclude that this integral diverges as $r_b\rightarrow\infty$. In order to 
have a finite result for the entanglement entropy  we must separate the divergent part from this expression as follows:
\begin{equation}\label{AAA}
 \mathcal{A}=\mathcal{A}_{\text{div}}+\mathcal{A}_{\text{fin}},
\end{equation}
 where the divergent part of this area can be defined in terms of the UV cut-off $r_b$ as:
\begin{equation}\label{m5-a-div}
 \mathcal{A}_{\text{div}}=L^4\int_{0}^{r_b}dr\frac{1}{f(r)}\sqrt{\frac{f^5(r_a)}{f^5(r_a)-1}}.
\end{equation}
From the equation for the area $\mathcal{A}$ (eq.$\ref{AAA}$)  and from the RT-formula (eq.\ref{s0}) we can schematically write down
the entanglement entropy as follows: 
\begin{equation}\label{ee5}
 S=S_{\text{div}}+\frac{L^4R_5}{4G_N^7}\,\,s\left(\frac{l}{R_5}\right),
\end{equation}
where the dimensionless function $s(l/R_5)$ is the finite part  of the entanglement entropy.  It is not always possible to 
express this dimensionless entropy $s$ as a function of $l$, however we can put this entropy in terms of $r_a$ and $r_b$ which follows 
from eqs.(\ref{m5-a}) and (\ref{m5-a-div}): 
\begin{equation}
s(r_a)=\int_{r_a}^{r_b}\frac{dr}{R_2}\frac{1}{f(r)}\sqrt{\frac{f^5(r_a)}{f^5(r_a)-f^5(r)}}-\int_{0}^{r_b}\frac{dr}{R_2}\frac{1}{f(r)}\sqrt{\frac{f^5(r_a)}{f^5(r_a)-1}}. 
\end{equation}


\subsection{Near horizon limit: $AdS_7\times S^4$}

This limit geometry of the N M5-branes are obtained if we take the approximation $R_5>>r$ which  implies that $f\approx \frac{R_5}{r}$.
In this case the dual theory is the $D=6$ superconformal field theory.
So if calculate the width of the strip $l$ (eq.\ref{l5}) in this limit we will obtain: 
\begin{equation}\label{l5ads}
\frac{l}{R_5}=\frac{2R_5^{1/2}}{r_a^{1/2}}\int_1^{\frac{r_b}{r_a}}dx\frac{1}{\sqrt{x^3(x^5-1)}}=\frac{4R_5^{1/2}c_5}{r_a^{1/2}}\left[1+
\mathcal{O}\left(\frac{r_a^3}{r_b^3}\right)\right],
\end{equation}
where $c_5=\frac{\sqrt{\pi}\Gamma(3/5)}{\Gamma(1/10)}$. In what follows we are going to disregard the superior order 
$\mathcal{O}\left(\frac{r_a^3}{r_b^3}\right)$ in our calculations.

Next we compute the entanglement entropy in this background limit. This follows from the equation for the area $\mathcal{A}$ (eq.\ref{m5-a})
and from the RT-formula (eq.\ref{s0}):
\begin{equation}
 S_A=\frac{L^4r_a^2}{4G_N^7R_5}\int_1^{\frac{r_b}{r_a}}dx\sqrt{\frac{x^7}{x^5-1}}=\frac{L^4r_b^2}{20G_N^7R_5}-\frac{c_5L ^4r_a^2}{8G_N^7R_5}.
\end{equation}
Notice that the first term of this expression  contains the UV cut-off $r_b$ and that this term diverges as $r_b\rightarrow\infty$. This divergent term
can be also be obtained from eq.(\ref{m5-a-div}).

Now in order to find the entanglement entropy $S_A$  in terms of the width of the strip $l$, we eliminate the 
parameter $r_a$ from the last two equations, as a result we find:
\begin{equation}\label{ee5ads}
S_A=\frac{L^4r_b^2}{20G_N^7R_5}-\frac{32L^4c_5^5R_5^5}{G_N^7l^4}.
\end{equation}
This result agrees with the one found in \cite{Ryu:2006bv}. Then the dimensionless finite entanglement entropy is in this case:
\begin{equation}
 s(l/R_5)=-\frac{128c_5^5R_5^4}{ l^4}.
\end{equation}


\subsection{Flat-space limit}

In this case, we will find the entanglement entropy in the approximate almost-flat space geometry which results from 
taking the limit $R_5<<r$ in our computations. In this limit, we have  that  $f\approx1+\frac{R_5^3}{3r^3}$.
In this asymptotic limit, the dual field theory (when it exist) is a non-conformal nor
supersymmetric  $D=6$  theory with $SU(N)$ gauge group.
Then if we calculate the width of the strip (eq.\ref{l5}) in this background limit we will obtain:
\begin{equation}\label{flatl}
 \frac{l}{R_5}\approx\frac{2\sqrt{3/5}r_a^{5/2}}{R_5^{5/2}}\int_1^{\frac{r_b}{r_a}}dx\sqrt{\frac{x^3}{x^3-1}}\approx
 2\sqrt{3/5}\left(\frac{r_a}{R_5}\right)^{3/2}\frac{r_b}{R_5}.
\end{equation}

Then the entanglement entropy in this background limit follows from eq.(\ref{m5-a}) and from the RT-formula (eq.\ref{s0}):
\begin{equation}
S_A\approx \frac{\sqrt{3/5}L^4r_a^{5/2}}{4G_N^7R_5^{3/2}}\int_1^{\frac{r_b}{r_a}}dx\sqrt{\frac{x^3}{x^3-1}}
=\frac{\sqrt{3/5}L^4r_a^{3/2}r_b}{12G_N^7R_5^{3/2}}-\frac{\sqrt{3/5}c_5^*L^4r_a^{5/2}}{4G_N^7R_5^{3/2}},
\end{equation}
where $c_5^*=\frac{\sqrt{\pi}\Gamma(2/3)}{\Gamma(1/6)}$. Note that the first term of this expression correspond to
the divergent part of the entropy. This is because  this term diverges as the UV cut-off $r_b\rightarrow\infty$. Also the divergent 
part of this entropy can  be obtained from eq.(\ref{m5-a-div}).

In order to have the entanglement entropy $S_A$ in terms of the width of the strip $l$, we eliminate the parameter $r_a$ from these last couple 
of equations, as a result we find:
\begin{equation}
S_A=\frac{L^4l}{24G_N^7}-\frac{c_5^*L^4R_5}{2^{11/3}(3/5)^{1/3}G_N^7}\left(\frac{l}{r_b}\right)^{5/3}.
\end{equation}
Then the dimensionless finite entanglement entropy is:
\begin{equation}
 s(l/R_5)=-\frac{c_5^*}{2^{5/3}(3/5)^{1/3}}\left(\frac{l}{r_b}\right)^{5/3}.
\end{equation}
This novel result stills contains the UV cut-off $r_b$, however, it continues
to be a finite function since the ratio $\frac{l}{r_b}$ is finite according to (\ref{flatl}).

\subsection{M5-brane and asymptotic limits}

\begin{figure}[h]
 \centering
 \includegraphics[width=10cm,height=6cm]{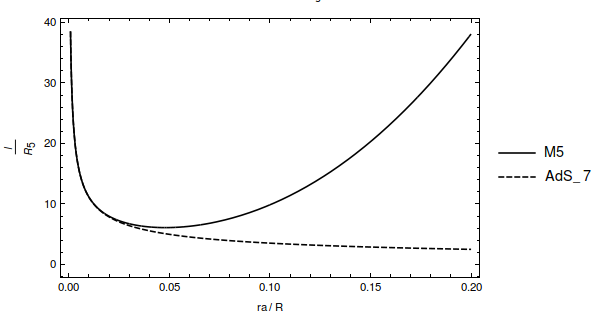}
 \caption{$l/R_5$ agaisnt $r_a/R_5$ for $r_a/R_5<0.2$ $r_b/R_5=10^3$.} 
 \label{fig51}
\end{figure}

In the M5-brane background, we can analyse numerically the entanglement entropy and compare this with its limit geometric background cases.
So first we  plot (see fig.\ref{fig51}) a comparison between the width of the strip $l/R_5$ for the cases of M2-brane and AdS$_7$.
We can observe  from this plot that the minimum  of $l/R_5$ ($l/R_5= 6.08$) for the M5-brane case is reached at $r_a/R_5=0.048$. These plots were
made at  $r_b/R_5=10^3$ and for the region $r_a/R_5<0.2$. 

\begin{figure}[h]
 \centering
 \includegraphics[width=9cm,height=7cm]{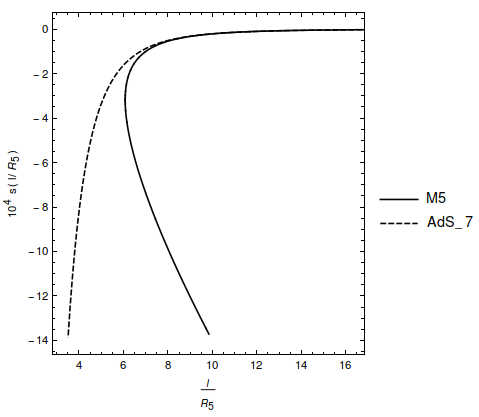}
 \caption{Entanglement entropy $s(l/R_5)$ for $r_b/R_5=10^3$.} 
 \label{fig52}
\end{figure}

Next we plot (see  fig.\ref{fig52}) a comparison between the dimensionless entropy $s(l/R_5)$ against $l/R_5$ for M5-brane and AdS$_7$
backgrounds cases. As we can see from this figure the dimensionless entropy for the M5-brane has two branches, 
the superior one corresponds to $r_a/R_5<0.048$ and the inferior one
corresponds to $r_a/R_5>0.048$. As we can notice  the superior brach coincides very well with the AdS$_7$ plot at the regime $r_a/R_5<0.048$. 

\begin{figure}[h]
 \centering
 \includegraphics[width=10cm,height=6cm]{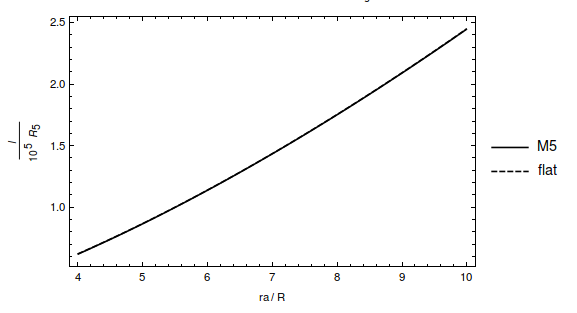}
 \caption{$l/R_5$ against $r_a/R_5$ for $r_a/R_5>4$ and $r_b/R_5=5\times10^3$.} 
 \label{fig53}
\end{figure}

Now we plot (see fig.\ref{fig53}) the comparison between the dimensionless entropy $s(l/R_5) $ for the M5-brane and almost-flat backgrounds cases. 
These plots were made in the regime $r_a/R_5>4$ and for the UV cut-off $r_b/R_5=5\times 10^3$.
As we can see from the figure the plots coincide very well in such regime.

\begin{figure}[h]
 \centering
 \includegraphics[width=10cm,height=6cm]{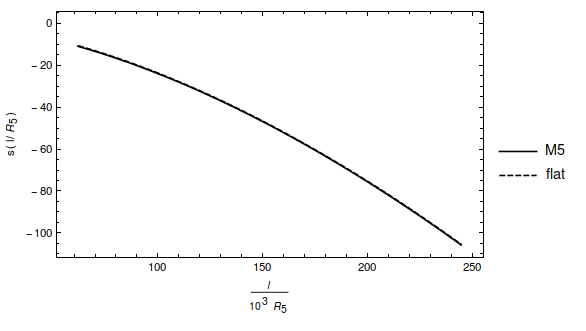}
 \caption{Entanglement entropy $s(l/R_5)$   for $r_a/R_5>4$ and $r_b/R_5=5\times 10^3$.} 
 \label{fig54}
\end{figure}

Next we plot (see fig.\ref{fig54}) a comparison between  the dimensionless entropy $s(l/R_5)$ against $l/R_5$ 
for the M5-brane and almost-flat backgrounds cases. These plot were made for the interval $r_a/R_5>4$ and 
for the  UV cut-off of $r_b/R_5=5\times 10^3$. As we can see these plots coincide very well in such a regime.


\section{Conclusions}

In this paper, in the context of Gauge/Gravity dualities, 
we have investigated some aspects of quantum entanglement 
in large N gauge theories that are dual to strings/M-theory 
in D3, M2-, M5-brane background spaces. Mainly, we have computed the quantum 
entanglement entropy of those theories applying the RT-formula  
for D3-, M2-, and M5-brane background spaces. This is done analytically 
for the asymptotic limit cases: AdS geometries, which results agree
with the literature, and the almost flat-space
geometries, that are novel results. Numerically, we have obtained 
entanglement entropies for
 the D3-, M2-, and M5-brane spaces and have compared them with their asymptotic
 geometric limis. 
 The entanglement
 entropies for these branes do not have in general a single valued behaviour, 
 instead all of these entropies have two branches, one going asymptotically (as
 the ratio $r_a/R$ decrease\footnote{$R$ represents here the radius of curvature of the branes 
 used in this work.})
 to the AdS behaviour and the other (as the $r_a/R$ increase) to the flat-space limits.
 
 We have also noticed that the choice of the UV cut-off value
plays a crucial role in the behaviour of 
the entanglement entropy function. For a big enough cut-off value,
the entanglement entropy function tends to have a decreasing behaviour
in all the backgrounds that we have studied.

It will be interesting to investigate the non-conformal dual theories that we mention in this work and also the entanglement  entropy for the case of non-extremal branes. In ref. \cite{Niarchos:2017cdz}, it has been suggested that the D3-brane case could be related to the open-closed string duality. This may offer a possible interpretation for the M2- and M5- cases. 

\acknowledgments We would like to thank interesting discussions with Carlos Nu\~nez. We also thank useful correspondence with Leo Pando-Zayas, Vasilis Niarchos, and Rohit Mishra. We acknowledge Conselho Nacional de Desenvolvimento Cient\'\i fico e Tecnol\'ogico (CNPq) - Brazilian agency - for financial support.  
				


\end{document}